\documentstyle[12pt]{article}
\begin{document}

\baselineskip 0.8cm

%======================================%
%<<<<<<<<<<<< TITLE PAGE >>>>>>>>>>>>>>%
%======================================%
\thispagestyle{empty}
\begin{center}
{\Large \bf Self-Tuning Dark Energy in Brane World Cosmology}
\end{center}

\vskip 1cm
\begin{center}
Kunihito Uzawa${}^1$ and Jiro Soda${}^2$\\
\vskip 0.5cm
${}^1${\it Graduate School of Human and Environment Studies, 
Kyoto University, Kyoto 606-8501, Japan}\\
${}^2${\it Department of Fundamental Science, FIHS, 
Kyoto University, Kyoto 606-8501, Japan}
\end{center}

%======================================%
%<<<<<<<<<<<<< ABSTRACT >>>>>>>>>>>>>>>% 
%======================================%
\vskip 4.5cm

\begin{center}
 {\Large Abstract}
\end{center}

Recently, the self-tuning mechanism of cancellation of vacuum
energy
 has been proposed in which our universe is a flat 3-brane in a
5-dimensional
 spacetime. In this letter, the self-tuning mechanism of dark energy is
 proposed by considering the cosmological matter in the brane world.
 In our model, the bulk scalar field takes the role of the dark energy
 and its value is slowly varying in time.
 The  claim is that even if the enormous amount of vacuum energy exists
 on the brane we can adjust the present value of the dark energy to
 be consistent with the current observations. In this self-tuning
 mechanism, the existence of the  constant of integration associated with
 the bulk scalar is crucial.

\newpage

The cosmological constant or vacuum energy  problem
\cite{wein}, \cite{car} has been
considered as a key issue from the view point of both particle
physics and observational cosmology for a long time.
From the particle physics point of view, the very large vacuum energy
 is problematical.
The vacuum energy density occurring after spontaneous symmetry
breaking at each cosmological stage is so large that the vacuum can
not be described by the Minkowski spacetime.
On the other hand, it is well known that our world is almost Minkowski
at least microscopically.
Consequently, we have to explain the reason why no vacuum energy is
left in the microscopic world.
From the observational cosmology point of view, the distance redshift
relation obtained by observing the Type Ia Supernova indicates
the accelerating universe\cite{obs}.
So in our universe, there exists the cosmological constant.
In fact, the cosmological observation suggests the existence of matter
 which violate the strong energy condition. This unknown component is now
 called  the dark energy.
Moreover, the combination of Type Ia Supernavae and the
first Doppler peak of Cosmic Microwave Background(CMB) anisotropy
indicates that the amount of the dark energy is about
$10^{-47}$GeV${}^4$\cite{sah}.
We must explain why such a value is taken in the universe.

Recently, N.Arkani-Hamed {\it et al.}\cite{ark} and, S.Kachru {\it et
 al.}\cite{kac} 
independently suggested a mechanism to realize Minkowski space-time
irrespective to the vacuum energy. They assumed that our world is
a 3-brane in 5-dimensional spacetime and the existence of the
scalar field in the bulk.
In the brane world scenario\cite{bra1}, \cite{bra2}, \cite{bra3}, \cite{bra4}
the standard model gauge and matter fields are assumed on the brane,
 while the gravity lives in the bulk. Due to this setup, the existence
 of the vacuum energy on the brane does not necessarily lead to the
 curved brane. Indeed, one can always find the
 Minkowski brane, although the bulk spacetime itself is curved.

This mechanism seems to  have solved, at least partially,
 the cosmological constant problem from the microscopic point of view.
 From the technical
point of view, the mechanism makes use of the fact that there always
exists a flat Friedmann universe provided the existence of the scalar field.
 Of course, the  spatial coordinate perpendicular to the brane world
 takes the role of the ``time'' in this argument.
 Anyway, it heavily relies on the specific slicing of
 the 5-dimensional space-time.
 On the other hand, the cosmological brane world can be regarded as an
appropriate timelike hypersuface in some 5-dimensional bulk space-time.
 Hence, it is not clear if this is the solution of the cosmological
 constant problem in the cosmology.
More precisely, we need the tiny amount of the dark energy compared to
 the energy scale of the electro-weak phase transition, not the vanishing
 cosmological constant, from the cosmological perspective.

 The purpose of this letter is to show that the action that resolve 
 the microscopic
 cosmological constant problem naturally leads to the
 ``dark energy'' component
 recently discussed in the astrophysics community \cite{obs}.

The situation we envisage is as follows: The effective 
4-dimensional cosmological constant is
expressed by the difference between brane tension and bulk cosmological
constant in the Randall\&Sundrum model\cite{ran}.   
 We assume that those will be canceled by the supersymmetry in the very
early Universe\cite{cve}. 
But the brane tension is slightly shifted 
 after electro-weak spontaneous symmetry breaking and therefore the
 effective cosmological constant no longer vanishes.
However, in the brane model, the self-tuning mechanism will work and the
observed dark energy could be explained.  
We will show this mechanism below.

Our starting point is 3-brane embedded in the 5-dimensional spacetime 
and the perfect fluid on the brane, 
We consider the system that the scalar field couples only with the  
brane tension.  This model is described by the following low energy
effective action: 

%============< EQUATION >==============%
%
\begin{eqnarray}
S &=& \frac{1}{2\kappa^2}\int d^5x\sqrt{-g^{(5)}}\:[R^{(5)}+2\lambda(\phi)]
  + \int d^4x\sqrt{-g}\:{\cal L}_m \nonumber\\
  & & - \int d^5x\sqrt{-g^{(5)}}\:\frac{1}{2}\left(\nabla\phi\right)^2
      - \int d^4x\sqrt{-g}\:f(\phi)\:\sigma,
\end{eqnarray}
%======================================%
where $\kappa$ is five dimensional gravitational constant,  
$R^{(5)}$ is five dimensional Ricci scalar,  
$\lambda$ is bulk cosmological constant which is a function of a 
scalar field $\phi$,
$g_{MN}$ is  5-dimensional metric, $f(\phi)$ is a scalar potential,
$\sigma$ is surface brane tension, and 
${\cal L}_m$ is the matter sector of the Lagrangian respectively.
We consider the following form for the five dimensional metric:  

%============< EQUATION >==============%
%
\begin{equation}
ds^2 = e^{\beta(y, t)}(-dt^2+dy^2) 
     + e^{\alpha(y, t)}\delta_{ij}dx^i dx^j,
\end{equation}
%======================================%
where $(i, j)=(1, 2, 3)$, and $y$ denotes the direction of extra 
dimension and 3-brane is located at $y=0$.
The energy-momentum tensor of the perfect fluid on the brane is

%============< EQUATION >==============%
%
\begin{equation}
{T^M}_N = diag (-\rho,\; p,\; p,\;p, 0)\:\exp{(-\beta)}\delta(y),
        \label{eq:energy}
\end{equation}
%======================================%
where $(M, N)=(0, 1, 2, 3, 4)$, $\rho$ and $p$ are the energy density 
and the pressure of the matter respectively. 
Moreover we impose the $Z_2$ symmetry under the transformation
$y\rightarrow -y$ for the spacetime.
This assumption is not essential for our analysis.
Once this metric and energy-momentum 
tensor are given, one can write easily the 
gravitational field equations and scalar field equation:

%============< EQUATION >==============%
%
\begin{eqnarray}
& &\dot{\alpha} + \dot{\alpha}\:\dot{\beta} - {\alpha}''
                  -2{{\alpha}'}^2 + {\alpha}'{\beta}'\nonumber\\
                & & \hspace{1cm}= \frac{1}{3}\lambda e^{2\beta} 
                 + \frac{1}{6}\kappa^2\left\{{\dot{\phi}}^2 + {\phi'}^2
                 + 2\left(\:f(\phi)\sigma+\rho\right)\delta(y) \right\},
                 \label{eq:00}\\
& &- 2\ddot{\alpha} - 3{\dot{\alpha}}^2 -\ddot{\beta} 
               + 2{\alpha}'' + 3 {{\alpha}'}^2 + {\beta}''\nonumber\\
             & & \hspace{1cm} = \lambda e^{2\beta} 
               + \frac{1}{2}\kappa^2\left\{{\dot{\phi}}^2 + {\phi'}^2
               + 2\left(\:-f(\phi)\sigma+p\right)\delta(y) \right\},
                  \label{eq:ij}\\        
& &- \ddot{\alpha} - 2{\dot{\alpha}}^2
                + \dot{\alpha}\dot{\beta} + {{\alpha}'}^2 
                + {\alpha}'{\beta}'
             = \frac{1}{3}\lambda e^{2\beta} 
                + \frac{1}{6}\kappa^2\left({\dot{\phi}}^2 + {\phi'}^2\right),
                  \label{eq:55}\\
& & {\beta}'\dot{\alpha} + {\alpha}'\dot{\beta} 
                  - {\dot{\alpha}}' - \dot{\alpha}\:{\alpha}'
                 = \frac{1}{3}\kappa^2\dot{\phi}\phi',
   \label{eq:051}\\
& & \kappa^{-2}\lambda'(\phi) + e^{-2\beta}\left(\ddot{\phi}
 -3\dot{\alpha}\dot{\phi}+{\phi}''+2\alpha'\phi'\right)
  -e^{-2\beta_0}f'(\phi_0)\sigma\delta(y)=0,
        \label{eq:dil}
\end{eqnarray}
%======================================%
where a prime stands for a derivative with respect to $y$, and a dot
denotes a derivative with respect to $t$.
Although $\alpha$, $\beta$ and $\phi$ are continuous in a neighborhood of the
 brane, their derivatives at $y=0$ are in general discontinuous
 across the brane. 
Then using gravitational field equations, we can obtain the junction
conditions

%============< EQUATION >==============%
%
\begin{eqnarray}
& &\alpha_1(t)\equiv\lim_{y\rightarrow 0+} 
                \frac{\partial\alpha(y, t)}{\partial y}
            = -\frac{1}{6}{\kappa}^2\left\{f(\phi_0)\sigma+\rho\right\}
              \:e^{\beta_0(t)},
        \label{eq:ja}\\
& &\beta_1(t)= \frac{1}{6}{\kappa}^2 \left\{-f(\phi_0)\sigma+2\rho + 3p\right\}
             e^{\beta_0(t)},
        \label{eq:jb}\\
& &\phi_1(t)= \frac{1}{2}\frac{d\:f(\phi_0)}{d\phi}\:\sigma 
     e^{\beta_0(t)},
        \label{eq:jp}
\end{eqnarray}
%======================================%
where $\beta_0(t)\equiv\beta\left(0, t\right) $ {\it etc}.
We note that to solve the junction condition with equation of state is 
equivalent to solve the gravitational field equation. 

As we consider the cosmological evolution on the brane, it is useful
to expand 
$\alpha$, $\beta$, and $\phi$ in power series with respect to $y$ at a
neighborhood of the brane\cite{flan}.
If we substitute these expansions into
each components of gravitational field equations, we get once again the
junction conditions (\ref{eq:ja})$\sim$(\ref{eq:jp}) on the brane.
By inserting the junction conditions (\ref{eq:ja})$\sim$(\ref{eq:jp}) into 
the equation (\ref{eq:051}), and picking up to order $y$, 
 the energy conservation law $\dot{\rho}+3(\rho+p)\dot{\alpha_0}=0$ 
is obtained.  In the same way, the Freidmann equation can also be
deduced from eq.(\ref{eq:55}) after the rescaling 
$d\hat{t} = \exp(\beta_0)dt$ \cite{flan}: 

%============< EQUATION >==============%
%
\begin{eqnarray}
& &H^2= -\frac{1}{6}\lambda(\phi_0)
 +\frac{1}{36}\kappa^4\:\left\{f(\phi_0)\right\}^2\sigma^2   
 +\frac{\kappa^4}{36}\rho\left\{2f(\phi_0)\sigma+\rho\right\}
\nonumber\\
& &\hspace{2cm}-e^{-4\alpha_0}\int d\alpha_0 
\left[e^{4\alpha_0}
\frac{d}{d\alpha_0}\left\{-\frac{1}{6}\lambda(\phi_0)
+\frac{1}{36}\kappa^4\:\left\{f(\phi_0)\right\}^2\sigma^2\right\}\right.
\nonumber\\
& &\hspace{2cm}\left.+\frac{\kappa^4}{18}\sigma\rho\frac{d}{d\alpha_0}f(\phi_0)
-\frac{\kappa^2}{3}\left\{\left(\frac{d\phi_0}{d\hat{t}}\right)
   ^2+\sigma^2\left(\frac{d\:f(\phi_0)}{d\phi}\right)^2\right\} \right],
\nonumber\\
           \label{eq:Fried2}
\end{eqnarray}
%======================================%
where the Hubble parameter is defined by $H = (d\alpha_0/d\hat{t})$. 
If the scalar field takes the constant value, this equation will reduce
to that obtained by Flanagan {\it et.al.}\cite{flan}.
The term 
$-\lambda(\phi_0)/6
 +\kappa^4\:\left\{f(\phi_0)\right\}^2\sigma^2/36$ on the right hand
 side is supposed to be canceled by supersymmetry\cite{cve}.
In addition, if the low energy condition $f\sigma\gg \rho$ is satisfied,
the conventional Friedmann equation with no cosmological constant is
recovered.

In the ordinary four dimensional cosmology, the cosmological constant 
can be read off from the Friedmann equation.
Thus effective dark energy $\Lambda_{eff}$ 
and effective gravitational constant $G_{eff}$ in our model are  
read off from the Freidmann equation (\ref{eq:Fried2}) respectively:

%============< EQUATION >==============%
%
\begin{eqnarray}
& &\Lambda_{eff} =-\frac{1}{6}\lambda(\phi_0) 
+\frac{1}{36}\kappa^4\:\left\{f(\phi_0)\right\}^2\sigma^2\nonumber\\
& &\hspace{2cm}-e^{-4\alpha_0}\int d\alpha_0 \left[e^{4\alpha_0}
\frac{d}{d\alpha_0}\left\{-\frac{1}{6}\lambda(\phi_0)
+\frac{1}{36}\kappa^4\:\left\{f(\phi_0)\right\}^2\sigma^2\right\}\right.
\nonumber\\
& &\hspace{2cm}\left. 
-\frac{\kappa^2}{3}\left\{\left(\frac{d\phi_0}{d\hat{t}}\right)
   ^2+\sigma^2\left(\frac{d\:f(\phi_0)}{d\phi}\right)^2\right\}\right],
\label{eq:effcos}\\
& &G_{eff} =  \frac{\kappa^4}{18}\sigma \left\{f(\phi_0)
-\frac{e^{-4\alpha_0}}{\rho}\int d\alpha_0 e^{4\alpha_0}
\rho\frac{d}{d\alpha_0}f(\phi_0) \right\}.
\label{eq:effgra}
\end{eqnarray}
%======================================%
The behavior of $\Lambda_{eff}$, $G_{eff}$ is found if we know 
$\lambda(\phi_0)$, $f(\phi_0)$, $\alpha_0$, and $\phi_0$. 
So, it is necessary
for us to solve the field equations to derive the dynamics of 
dark energy .  
However, it is very difficult to solve the
junction condition with matter on the brane.   
So at first, we discuss an approach to solve the
field equations. 
The point is that the brane world cosmology can be constructed by
cutting the 5-dimensional static solution along the suitable slicing and
gluing two copies of remaining spacetime. 
The jump of the extrinsic curvature along the slicing should be
equated with the matter localized on the brane.  
  Therefore the cosmological solution with matter can be obtained by
 a suitable slicing. In order to find it, we have to seek a
coordinate transformation leading to the slicing which determines the
matter on the brane. The coordinate transformations will be found by
imposing the equation of state.
Now we use this transformation method from the static brane solution to the
cosmological brane solution\cite{koya}. 
First of all, we look for the static solution with the following metric:
%============< EQUATION >==============%
%
\begin{equation}
ds^2 = \exp\left\{2\alpha(z)\right\}
        \left(-d\tau^2 + dz^2 + \delta_{ij}dx^i\:dx^j\right).   
        \label{eq:stm}
\end{equation}
%======================================%
Here we will take the simple ansatz to solve gravitational field equations.
That is, we assume that $\phi$ is related to $\alpha$ as  
$\phi = A\alpha + B$, where $A$ and $B$ are constants. 
We note that the constant $B$ will control the
smallness of four dimensional dark energy value as we mention 
later on.
The form of $\lambda$ is assumed as
 $\lambda(\phi)=\Lambda\:e^{b\phi_0}$, where $\Lambda$ is constant. 
The equation for $\alpha$ can be easily obtained as  

%============< EQUATION >==============%
%
\begin{equation}
\alpha = - \left(1-\frac{3b^2}{4\kappa^2}\right)^{-1}\:
        \ln \biggl|\frac{z}{l}\biggr|,
  \label{eq:alpha}
\end{equation}
%======================================%
where $l$ is a constant of integration and $A$ is determined as 
$A = -3b/2\kappa^2$. It should be noted that $b\rightarrow 0$ corresponds to 
Anti-de Sitter spacetime (AdS) with the curvature scale $l$.
The constant $B$ is also determined as:

%============< EQUATION >==============%
%
\begin{equation}
e^{bB} = \frac{6}{\Lambda\:l^2}\left(1-\frac{3b^2}{16\kappa^2}\right)
\left(1-\frac{3b^2}{4\kappa^2}\right)^{-2}.
\end{equation}
%======================================%
The Ricci scalar on this background can be calculated as

%============< EQUATION >==============%
%
\begin{eqnarray}
R^{(5)} &=& -\biggl|\frac{z}{l}\biggr|^{2\left(1
-\frac{3b^2}{4\kappa^2}\right)^{-1}-2}\left\{\frac{20-\frac{6b^2}{\kappa^2}}
{l^2\left(1-\frac{3b^2}{4\kappa^2}\right)^2}\right\}.
\end{eqnarray}
%======================================%
Thus this background has singularity at $z\rightarrow\infty$ for
$(3b^2/4\kappa^2)<1$.  
In order to make the scenario complete, the appropriate boundary
condition must be imposed at the singularity. This defect is common in
the brane world scenario and is believed to be resolved in the string
theory framework\cite{gub}.
Here, in order to seek an appropriate slicing, we make a coordinate
transformation $(\tau, z)$ to $(u, v)$ system:
$\tau = l\left\{h(u) + g(v)\right\}, \hspace{0.2cm}
z = l\left\{h(u)-g(v)\right\}$, where $h$ and $g$ are 
 functions of $u$ and $v$, respectively.
Next, we assumed the following relation between $(u, v)$ and $(t, y)$
as $u=(t - y)/l$ and $v=(t + y)/l$.
The $(u, v)$ is null coordinates in the new coordinate system. The function
$h$ and $g$ are transformation functions between null 
coordinate on the static brane and that on the cosmological brane. 
So line element is written by new coordinate $(t ,y)$ as 
follows

%============< EQUATION >==============%
%
\begin{equation}
ds^2 = \frac{4h'(u)g'(v)}
       {\left[
       \left\{h(u) - g(v)\right\}\right]^{2\left(1-\frac{3b^2}{4\kappa^2}
       \right)^{-1}}}\left(-dt^2 + dy^2\right)
      + \frac{1}{\left[
       \left\{h(u) - g(v)\right\}\right]^{2\left(1-\frac{3b^2}{4\kappa^2}
       \right)^{-1}}}\:\delta_{ij}\:dx^i\:dx^j, 
   \label{eq:le}
\end{equation}
%======================================%
where $h'$ and $g'$ denote $dh(u)/du$ and  
$dg(v)/dv$ respectively. 

In principle, the functions $h(u)$ and $g(v)$ are determined by the
junction conditions with the appropriate equation of state.
This process is very difficult to perform without any approximation. 
In ref.\cite{koya}, the low energy expansion is used to obtain the
explicit form of $h$ and $g$. Fortunately, for our purpose, it is not
necessary to know the exact solutions. The qualitative behavior of the
cosmological solution can be inferred from the relation to the static
solution through the coordinate transformation.  
The expression for $a(t)$ is given by
%============< EQUATION >==============%
%
\begin{equation}
a(t) \equiv e^{\alpha_0}
         = \left[
     \left\{h\left(\frac{t}{l}\right) - g\left(\frac{t}{l}
     \right)\right\}\right]^{-\left(1-\frac{3b^2}{4\kappa^2}\right)^{-1}}.
\end{equation}
%======================================%
where, we fixed the gauge freedom so that $t$ is cosmological time 
(In short, $e^{2\beta_0}=1$).  
Thus, we find that $z\rightarrow\infty$ corresponds to initial 
singularity while 
$z\rightarrow 0$ corresponds to future infinity. 
We note that time $t$ dependence of $a$ is described through a change
of $z$. 
And, using scale factor, $\phi_0$ is expressed 
 as $\phi_0(t) =  \ln a(t)^{-\frac{3b}{2\kappa^2}} + B$ from the ansatz.
$\phi=A\alpha+B$. 
Here we shall specify the scalar potential $f(\phi_0)$ in order to know the
 dynamics of $\Lambda_{eff}$ and $G_{eff}$. 
The equation between $\alpha_1$ and $\phi_1$ is given by the ansatz as 
$\phi_1=-\frac{3b}{2\kappa^2}\alpha_1$.
While the relation
$\phi_1=-\frac{3}{\kappa^2}\frac{df(\phi_0)}{f(\phi_0)}\alpha_1$ is 
obtained by the junction conditions (\ref{eq:ja}) and (\ref{eq:jp})
because we consider the low energy case 
$f\sigma\gg\rho$. The quantitative justification of this relation will
be  discussed later.
Thus we get the form $f(\phi_0) = \:e^{b\phi_0/2}$ as 
the scalar field potential.
Therefore, $f(\phi)$ is allowed  
to have only  the form of exponential type under $Z_2$ symmetry
 to be consistent with junction 
conditions on the brane (\ref{eq:ja}) and (\ref{eq:jp}). 
Using the eqs.(\ref{eq:effcos}) and (\ref{eq:effgra}) 
and this potential, we can write the effective 
dark energy and gravitational constant as follows

%============< EQUATION >==============%
%
\begin{eqnarray}
\Lambda_{eff} &=& 
\left(\frac{1}{36}\kappa^4\sigma^2-\frac{1}{6}\Lambda-\frac{1}{192}
\kappa^2b^2\sigma^2\right)\left(1-\frac{3b^2}{8\kappa^2}\right)^{-1}
a^{-\frac{3b^2}{2\kappa^2}}e^{bB}
-\frac{3b^2}{4\kappa^2}a^{-4}\int dt\:a\:{\dot{a}}^3,
\label{eq:c0}\\
G_{eff}&=&\frac{1}{18}\kappa^4\sigma\:
\left\{1+\left(\frac{3b^2}{4\kappa^2}\right)
\left(1-3w-\frac{3b^2}{4\kappa^2}\right)^{-1}\right\}
 a^{-\frac{3b^2}{4\kappa^2}}\:e^{\frac{bB}{2}}, 
\label{eq:g0}
\end{eqnarray}
%======================================%
where we use the relation $\rho=e^{-3(1+w)\alpha_0}$
(This is derived by the combination of equation of 
state $p=w\rho$ and the energy conservation law).

The upper limit of the present rate of change of $G_{eff}$,   
 ${\dot{G}}_{eff}/G_{eff}$, 
 obtained by the Viking radar ranging and lunar laser ranging are given  
by ${\dot{G}}_{eff}/G_{eff}\le 8\times10^{-12}{\rm yr}^{-1}$ and  
${\dot{G}}_{eff}/G_{eff}\le(0.2\pm0.4)\times10^{-11}{\rm yr}^{-1}$,
 respectively \cite{hell},\cite{will}.
Therefore, the parameters are constrained as 
$\bigl|-3b^2/4\kappa^2\bigr|<(0.2\pm0.4)\times10^{-1}$. 

The second term in eq.(\ref{eq:c0}) will decay faster than $O(t^{-2})$ if
the scale factor has power law dependence on time. So its evolution 
at present is almost determined by the first term, which is found
 approximately ``constant'' from the constraint of experiment of the 
 rate of change of $G_{eff}$. 
Thus hereafter we neglect the second term in eq.(\ref{eq:c0}).

The ``dark'' energy we can directly measure is vacuum energy density 
$\rho_{\Lambda}$, which is defined by   
%============< EQUATION >==============%
%
\begin{eqnarray}
\rho_{\Lambda}\equiv\frac{3\Lambda_{eff}}{8\pi\:G_{eff}}
&\simeq&\frac{3\sigma}{16\pi}\left(1-X\right)
\left(\frac{1}{\Lambda\:l^2}\right)^{\frac{1}{2}},
\end{eqnarray}
%======================================%
where $X\equiv 6\Lambda\kappa^{-4}\sigma^{-2}$ and we used the relation 
$\kappa^{-2}b^2\ll 1$.
So the dark energy must be very slowly varying,
 which is ensured by the constraint of $G_{eff}$. 

In the Randall\&Sundrum model\cite{ran}, $\rho_{\Lambda}$ is zero 
(in other words $X=1$) by supersymmetric cancellation\cite{cve}. 
But in our system, the factor
 $1-X$ is slightly shifted from zero in terms of the spontaneous
symmetry breaking because $X$ is the ratio of brane tension between two
 cosmological phase.
Numerically, the resulting value is not sufficiently small to be
consistent with the observational constraint. However, there exists
 further freedom in our model.  
Recall that $l$ is the constant of integration and hence not yet
determined. Its value can be chosen to fit to the observation. This is
the cosmological self-tuning mechanism we have found.
Here we will list some numbers as an illustration.
The dark energy is expressed by 
 $\rho_{\Lambda}\simeq(1-X)\sigma e^{bB/2}$. 
We can write $1-X\sim
\delta\sigma/\sigma\sim{M_s}^{4}{M_{\sigma}}^{-4}$, where 
$M_{\sigma}$ and $M_s$ are the energy scale of $\sigma$ and 
symmetry breaking scale, respectively. If the symmetry
 breaking scale is taken as electro-weak scale $M_s\sim10^{2}\:$GeV and 
 $M_{\sigma}$ is 
set to be $M_{\sigma}\sim 10^{18}\:$GeV, then
$\rho_{\Lambda}\sim10^{9}({\rm GeV})^4 e^{bB/2}$.
On the other hand, the value of dark energy is
 $\rho_{\Lambda}\sim 10^{-47}$GeV${}^4$ supported 
by recent cosmological observation\cite{obs}. If the
value of $e^{bB/2}$ is self-tuned as $10^{-55}$, this result is
obtainable.
The effective gravitational
 constant becomes $G_{eff}\sim\kappa^4\sigma\:f\sim{M_{\sigma}}^{4}
{M_{\kappa}}^{-6}e^{bB/2}$.
If we take $M_{\kappa}\sim 10^9{\rm GeV}$, the correct value of the
Newton constant can be obtained.
We emphasize that if one set these energy scales, $1-X\sim 10^{-63}$ is 
derived quantity. On the other hand, $e^{bB/2}\sim
10^{-55}$ is selected by self tuning.  
Moreover, in this case 
$f\sigma\sim\sigma e^{bB/2}\sim(10^4{\rm GeV})^4\gg\rho$.
Thus, we have checked the low energy condition used in obtaining $f(\phi_0)$.

To conclude, we have proposed the model which exhibits the self-tuning of the
 dark energy in the brane world.
The difference between our model and N.Arkani-Hamed {\it et al.}\cite{ark},
S.Kachru {\it et al.}\cite{kac} model is inclusion of the cosmological matter
 in addition to the vacuum energy density on the brane.
In our model, the bulk scalar field couples directly with the vacuum energy
 on the brane but not with the cosmological matter. However, in the
 effective Friedmann equation, the scalar field couples with the
cosmological
 matter.  This is due to the intermediation by the gravity.
The small $\rho_{\Lambda}$ is allowed by the self-tuning.
In other words, it is always possible to find the solution
 with a value of dark energy
 measured by Today's observation.
 This is {\it not} the fine tuning of the parameters in the model
 but the self tuning mechanism in the same sense as the proposal by
 N.Arkani-Hamed {\it et al.}\cite{ark} and S.Kachru {\it et al.}\cite{kac}.

 Apparently, our model does not provide the complete solution of the
 cosmological constant problem in the observational cosmology.
 We have just shown that it is possible to find an observationally
 allowable solution. The situation is similar to the changing gravity
 solution explained in the review by Weinberg\cite{wein}. 
The final resolution of
the problem might be provided by the quantum gravity or anthropic principle.

\vskip 1cm

\noindent
Ackknowledgements.-
We are grateful to K. Koyama for helpful comments.

%======================================%
%<<<<<<<<<<<< REFERENCES >>>>>>>>>>>>>>%
%======================================%

\end{document}